\documentclass[runningheads]{llncs}
\usepackage[T1]{fontenc}
\usepackage{amsmath,amssymb,amsfonts}
\usepackage{graphicx}
\usepackage{xcolor}

\usepackage{multicol}
\usepackage{multirow}
\usepackage{placeins}

\usepackage{centernot}
\usepackage{stmaryrd}
\usepackage{tabularx}

\usepackage{makecell}

\newcolumntype{L}[1]{>{\raggedright\arraybackslash}p{#1}}
\newcolumntype{R}[1]{>{\raggedleft\arraybackslash}p{#1}}
\newcolumntype{C}[1]{>{\centering\arraybackslash}p{#1}}

\usepackage{transparent}
\usepackage{comment}

\newcommand{\C}{\mathbb C}
\newcommand{\R}{\mathbb R}

\newcommand{\Rd}{\mathbf{R}}

\newcommand{\CC}{\mathbf{c}}

\newcommand{\XX}{\mathbf{x}}  
                     
\newcommand{\ZZ}{\mathbf{z}}                       
\newcommand{\YY}{\mathbf{y}}

\newcommand{\HH}{\mathbf{h}}  
                         
\newcommand{\Ad}{\mathbf A}                     
 
\newcommand{\Fd}{\mathbf F} 
\newcommand{\Bd}{\mathbf B} 
\newcommand{\Id}{\mathbf I}
\newcommand{\Cd}{\mathbf C} 
\newcommand{\Ed}{\mathbf E}

\newcommand{\Fu}{\mathbf{F}_I} 
 
\newcommand{\Au}{\mathbf{A}_I}  
\newcommand{\Su}{\mathbf{S}_I}

\newcommand{\Yu}{\mathbf{y}_I}

\newcommand{\herm}{{\scriptstyle \boldsymbol{\mathsf{H}}}}
\newcommand{\trans}{{\scriptstyle \boldsymbol{\mathsf{T}}}}

\usepackage[percent]{overpic}
\usepackage{algorithm,algcompatible,amsmath}
\DeclareMathOperator{\diag}{diag}
\DeclareMathOperator*{\argmin}{arg\,min}
\algnewcommand\INPUT{\item[\textbf{Input:}]}%
\algnewcommand\PARAMETER{\item[\textbf{Parameters:}]}%
\algnewcommand\OUTPUT{\item[\textbf{Output:}]}%
\usepackage[
 colorlinks=false
]{hyperref}

\usepackage{makecell}
\usepackage{booktabs}

\usepackage{siunitx}
\usepackage{array}

\newcolumntype{L}[1]{>{\raggedright\let\newline\\\arraybackslash\hspace{0pt}}m{#1}}
\newcolumntype{C}[1]{>{\centering\let\newline\\\arraybackslash\hspace{0pt}}m{#1}}
\newcolumntype{R}[1]{>{\raggedleft\let\newline\\\arraybackslash\hspace{0pt}}m{#1}}

\begin{document}
\title{NoSENSE: Learned Unrolled Cardiac MRI Reconstruction Without Explicit Sensitivity Maps}
\titlerunning{NoSENSE: Learned Unrolled Cardiac MRI Reconstruction}
\author{Felix Frederik Zimmermann \and
Andreas Kofler}
\authorrunning{F.\ F.\ Zimmermann \and A.\ Kofler}
\institute{Physikalisch-Technische Bundesanstalt (PTB) Berlin, Germany
\email{felix.zimmermann@ptb.de}\\
}
\maketitle              
\begin{abstract}
We present a novel learned image reconstruction method for accelerated cardiac MRI with multiple receiver coils based on deep convolutional neural networks (CNNs) and algorithm unrolling. In contrast to many existing learned MR image reconstruction techniques that necessitate coil-sensitivity map (CSM) estimation as a distinct network component, our proposed approach avoids explicit CSM estimation. Instead, it implicitly captures and learns to exploit the inter-coil relationships of the images. Our method consists of a series of novel learned image and $k$-space blocks with shared latent information and adaptation to the acquisition parameters by feature-wise modulation (FiLM), as well as coil-wise data-consistency (DC) blocks.

Our method achieved PSNR values of $34.89$ and $35.56$ and SSIM values of $0.920$ and $0.942$ in the cine track and mapping track validation leaderboard of the MICCAI STACOM CMRxRecon Challenge, respectively, ranking 4th among different teams at the time of writing. 

Code is available at \url{https://github.com/fzimmermann89/CMRxRecon}
\end{abstract}
\section{Introduction}\label{subsec:intro}
Deep Learning-based approaches have recently been widely applied in the field of image reconstruction across different imaging modalities \cite{schlemper2017deep,adler2018learned,hammernik2018learning,aggarwal2018modl}. Recently, several image reconstruction challenges have been organized for different image reconstruction problems \cite{muckley2020state,knoll2020advancing,mccollough2017low}, and neural networks based on algorithm unrolling \cite{monga2021algorithm} are often among the winning teams.

Here, we present a new method developed for the MICCAI STACOM CMRxRecon Challenge, where the task is to reconstruct cardiac cine (and quantitative mapping) MR images from undersampled $k$-space measurements. In the following, we will mainly focus on the cine reconstruction. Our method is based on algorithm unrolling and utilizes both model-based and learned components. Unlike other recently published methods (e.g., \cite{sriram2020end,yiasemis2022recurrent}), our approach does not explicitly estimate coil sensitivity maps (CSM), and we argue that employing CSMs might potentially introduce model errors for the considered task.

\section{Methods}
\subsection{Theory and Motivation}\label{subsec:theory}
Let $\XX_{\mathrm{true}} \in \C^N$ be the vector representation of the unknown cardiac MR image with $N=N_x\cdot N_y \cdot N_z \cdot N_t$. The forward model considered in multi-coil MRI is given by 
\begin{equation}\label{eq:forward_model}
    \Yu:= \Au(\Cd)\XX_{\mathrm{true}} + \mathbf{e},
\end{equation}
where $\Au(\Cd): \C^N \rightarrow \C^M$ denotes the multi-coil MR operator with
\begin{equation}\label{eq:operator_mri}
    \Au(\Cd):= (\Id_{N_c} \otimes \Fu) \Cd, \quad \text{ with } \quad   \Cd^\herm \Cd= \Id_{N}
\end{equation}
where $\Cd = [\Cd_1, \ldots, \Cd_{N_c}]^\trans\in \C^{(N_c\cdot N) \times N}$ with $\Cd_c = \diag(\mathbf{s}_c) \in \C^{N \times N}$  denotes the operator of the $N_c$  stacked CSMs, which are represented by diagonal operators. Here, $\otimes$ represents the Kronecker product, $\Fu:=\Su \Fd$ with $\Fd$ being a frame-wise 2D Fourier operator, $\Su$ a binary sub-sampling mask that collects a subset $I \subset J=\{1,\ldots, N_x\cdot N_y\}$ of the $k$-space coefficients of a 2D image, $\Id_{N_c}$ and $\Id_{N}$ are the $N_c$- and $N$-dimensional identity operators, respectively, and $\mathbf{e}$ is complex-valued Gaussian noise.
Additionally, let $\Ad$ be defined as in \eqref{eq:forward_model} but with $I=J$, i.e.\ it models a full acquisition with no acceleration, and thus, $\Su = \Id_N$.
The operator $\Au$ depends on $\Cd$ which is unknown and must be estimated with appropriate methods \cite{uecker2014espirit}. Once $\Cd$ is fixed, one can consider variational problems of the form
\begin{equation}\label{eq:reg_problem}
    \underset{\XX }{\min}\, \mathcal{F}_{\lambda}(\XX), \quad  \mathcal{F}_{\lambda}(\XX):=\frac{1}{2}\|\Au(\Cd) \XX - \Yu\|_2^2 + \lambda\, \mathcal{R}(\XX),
\end{equation}
where $\mathcal{R}(\XX)$ encodes the regularizing properties that can be learned from data and $\lambda>0$. From \eqref{eq:reg_problem}, different reconstruction methods based on algorithm unrolling \cite{monga2021algorithm} can be derived. For example, by identifying the learned components as learned proximal operators \cite{cheng2019model}, learned sparsifying transforms \cite{kofler2022convolutional}, learned filter transforms with learned potential functions \cite{hammernik2018learning} or learned denoisers/artifact reduction methods \cite{schlemper2017deep,aggarwal2018modl}. 
Generally, these methods alternate between the application of learned blocks and model-based blocks, which use information on the forward and adjoint operators $\Au$ and $\Au^\herm$. For multi-coil acquisitions, the latter blocks often require solving inner minimization problems \cite{aggarwal2018modl,schlemper2017deep,duan2019vs}. Last, instead of first estimating the operator $\Cd$ and fixing it when deriving a reconstruction network from \eqref{eq:reg_problem}, recent methods often estimate CSMs using special blocks within a learned reconstruction \cite{sriram2020end}, \cite{yiasemis2022recurrent}.

In this work, we adopt a strategy based on the structure of the specific problem formulation and task posed by the CMRxRecon Challenge. Unlike other methods, our approach does not require a prior estimate of the CSMs or their refinement within the network architecture. We illustrate that unrolled methods derived from a variational problem as in \eqref{eq:reg_problem} may introduce systematic errors that the learned components must compensate for, rendering them sub-optimal.

\subsection{CMRxRecon Dataset}\label{subsec:dataset}
We briefly describe the given dataset (see \cite{cmrxrecon} for more details) and the associated challenge task motivating our proposed approach. 
Let the multi-coil $k$-space data acquired according to \eqref{eq:forward_model} be of the form $\Yu = [\Yu^1, \ldots, \Yu^{N_c}]^\trans$. The provided dataset  consists of undersampled  $k$-space data of short- and long-axis (SAX/LAX) images, retrospectively undersampled with different acceleration factors $R\in \{4,8,10\}$ (not counting the 24 central ACS lines that are always sampled) from fully-sampled data and root sum of squares (RSS) reconstructions obtained from the latter, i.e.\ $\mathcal{D}=\big\{(\Yu, \Su, \YY, \XX_{\mathrm{RSS}})\}$, where
\begin{align}
    \Yu = \Su \YY,\quad \YY:=\Ad(\Cd_{\mathrm{true}})\XX_{\mathrm{true}} + \mathbf{e}, \quad  \XX_{\mathrm{RSS}} = \Big(\sum_{c=1}^{N_c} | \Fu^\herm \Yu^c|^2\Big)^{1/2}.
\end{align}
 Importantly, we note that, first, we do not have access to the pairs $(\Cd_{\mathrm{true}},\XX_{\mathrm{true}})$  that generate the complete $k$ space data $\YY$ according to the forward model as in \eqref{eq:forward_model}. Secondly, due to the retrospective undersampling, the undersampled data as well as the targets are subject to the same realization of the random noise $\mathbf{e}$. Finally, the task of the challenge is to obtain an estimate of $\XX_{\mathrm{RSS}}$ from $\Yu$, rather than finding an estimate of $\XX_{\mathrm{true}}$. This has an important consequence as illustrated in the following.
 
 Let $\Bd: V \rightarrow W$ be a linear operator between Hilbert spaces $V$ and $W$ (possibly the forward model in \eqref{eq:operator_mri}), and $\{\Rd_{\lambda}\}_{\lambda>0}$ with $\Rd_{\lambda}:W \rightarrow V$ be a family of continuous operators. Then, $\{\Rd_{\lambda}\}_{\lambda>0}$ is a regularization of the Moore-Penrose generalized inverse $\Bd^\dagger$, if $\Rd_{\lambda} \rightarrow \Bd^\dagger$ for $\lambda \rightarrow 0$ pointwise on the domain of $\Bd^\dagger$. In particular, for the fully-sampled case of $\Ad$ in \eqref{eq:forward_model}, we have
 \begin{equation}\label{eq:sol_vs_rss}  \XX^{\ast}:=\argmin_{\XX}\,\mathcal{F}_{\lambda}(\XX) \overset{\lambda \rightarrow 0 }{\longrightarrow} \Ad^\dagger \YY = \Ad^\herm \YY = \sum_{c=1}^{N_c} \Cd_c^\herm \Fu^\herm \Yu^c,
\end{equation}
 where we used that $\Ad^\dagger:=(\Ad^\herm \Ad)^{-1} \Ad^\herm = \Ad^\herm $  due  the normalization of $\Cd$ in \eqref{eq:operator_mri}. From \eqref{eq:sol_vs_rss}, we observe that even in the fully-sampled case,  $|\XX^{\ast}| \overset{\lambda \rightarrow 0 }{\longarrownot\longrightarrow} | \XX_{\mathrm{RSS}}|$ due to the CSMs.  Specifically, this implies that any reconstruction network designed to approximate a solution to problem \eqref{eq:reg_problem} inherently introduces an error that the learned component must counteract. This observation motivates the construction of a network that addresses a different problem than stated in \eqref{eq:reg_problem}. Rather than learning to obtain an estimation of the noise-free complex-valued ground-truth image (and subsequently utilizing its magnitude as an estimate for the RSS-reconstruction), it seems a more promising approach to learn to estimate fully-sampled (potentially noisy) $k$-space data, obtain the respective coil-images by a simple Fourier transform, and finally perform an RSS-reconstruction. Indeed, there have been methods proposed that estimate $k$-space data instead of the underlying image \cite{yiasemis2022recurrent}; nevertheless, these methods lack an explanation for the rationale behind this preference.
 
\subsection{ Proposed Approach }\label{subsec:problem_formulation}
We propose a learned method that, instead of estimating $\XX_{\mathrm{true}}$ by deriving a reconstruction scheme from \eqref{eq:reg_problem}, directly estimates the fully-sampled multi-coil $k$-space data $\tilde{\YY}:=[\tilde{\YY}^1, \ldots, \tilde{\YY}^{N_c}]^\trans$. Given $\tilde{\YY}$, an RSS can be performed to obtain an estimate of $\XX_{\mathrm{RSS}}$. Hence,  a suitable problem formulation is
\begin{equation}\label{eq:reg_problem_k}
    \underset{\YY=(\YY^1, \ldots, \YY^{N_c}) }{\min}\, \mathcal{G}(\YY), \quad \,  \mathcal{G}(\YY):= \mathcal{S}_{\theta}(\YY) \,\, \text{ such that } \,\, \Su \YY = \Yu,
\end{equation}
where $\mathcal{S}_{\theta}(\YY)$ encodes the regularization imposed on the sought $k$-space data. Comparing \eqref{eq:reg_problem_k} and \eqref{eq:reg_problem}, we observe two key differences: first, the problem is now formulated in $k$-space rather than in image space. Second, instead of only requiring approximate data-consistency between the image and the acquired $k$-space data, it might be desirable to obtain hard data consistency, as suggested by the constraint in \eqref{eq:reg_problem_k} in the absence of noise.

In line with related approaches \cite{schlemper2017deep}, \cite{sriram2020end}, we propose to construct a cascade that alternates between the application of CNN-blocks as well as data-consistency layers. Instead of relying on CSM estimation, we work directly with the coil-weighted $k$-space/ image data, i.e.\ each vector in the following consists of $N_c$ components. Let $\Ed:= (\Id_{N_c} \otimes \Fd)$ denote the operator that performs a frame-wise 2D Fourier transformation on each coil-image, and $\Ed^\herm$ its adjoint. We unroll $T$ iterations of the following scheme for $k=0,\ldots,T-1$:
\begin{eqnarray}\label{eq:scheme}
\tilde{\ZZ}_k, &\tilde{\HH}_{k+1} &:=\mathrm{X}_{\theta}\big(\Ed^\herm \YY_{k}, \tilde{\HH}_{k}, \mathbf{c}_k\big), \label{eq:img_cnn}\\
\ZZ_k, &\HH_{k+1} &:= \mathrm{Y}_{\theta} \big(\Ed\,\tilde{\ZZ}_{k},\HH_{k}, \mathbf{y}_I,\mathbf{c}_k\big), \label{eq:kspace_cnn}\\
    &\YY_{k+1} &:=\big[\underset{\YY^1 }{\argmin}\, \mathcal{G}_{\lambda^1_k, \ZZ_{k}^1}(\YY^1), \ldots, \underset{\YY^{N_c} }{\argmin}\, \mathcal{G}_{\lambda^{N_c}_k,\ZZ_{k}^{N_c}}(\YY^{N_c}) \big]^\trans , \label{eq:scheme_min_problem}\\
    &\tilde{\XX}_{\mathrm{RSS}}&:= \mathrm{RSS}(\YY_T)
\end{eqnarray}
where $\mathrm{X}_{\theta}$ and $\mathrm{Y}_{\theta}$ denote U-Net \cite{ronneberger2015u} blocks operating in image- and $k$-space, respectively \cite{kiki}. In both, information from the latent space is passed as hidden variables to the next iteration as $\HH_{k+1}$ and $\tilde{\HH}_{k+1}$, respectively, inspired by the recent success of recurrent networks \cite{yiasemis2022recurrent} for MR reconstruction.
Moreover, the blocks use information about the acquisition, i.e., the axis (SAX or LAX), the slice / view, the current iteration $k$, and the acceleration factor $R$, which are encoded in one hot and transformed into $\mathbf{c_k}$ by a fully connected network (MLP). In addition, $\mathrm{Y}_\theta$ utilizes the acquired undersampled $\YY_I$ data as additional input in each iteration.

The functionals in \eqref{eq:scheme_min_problem} take the form
\begin{equation}\label{eq:reg_functional_coil}
    \mathcal{G}_{\lambda^c_k, \ZZ_{k}^c}(\YY^c) = \frac{1}{2}\big\| \Su \YY^c - \Yu^c\big\|_2^2 + \frac{\lambda^c_k}{2}\big\|   \YY^c - \ZZ_{k}^c\|_2^2
\end{equation}
and separately regularize each coil-weighted $k$-space data. Note that the CNNs in \eqref{eq:scheme}, in contrast, jointly utilize all coil information. 
For each DC block and each coil $c$, we use a learned linear function with bias from $\mathbf{c}_k$ to $\lambda^c_k$. In contrast to \eqref{eq:reg_problem_k}, no hard data-consistency is enforced in \eqref{eq:reg_functional_coil}. In preliminary experiments, we experimented with hard data-consistency constraints by projecting the intermediate $k$-space outputs onto the kernel of the sub-sampling operator $\Su$ in the spirit of \cite{schwab2019deep}. We observed a more stable training when employing a soft data-consistency  in \eqref{eq:scheme_min_problem}, i.e., solving \eqref{eq:reg_functional_coil} as \cite{ravishankar2010mr}
\begin{equation}
        \YY_{k+1}^c = \mathbf{S}_I \left( \frac{1}{1+\lambda^c_k} \YY^c_I + \frac{\lambda^c_k}{1+\lambda^c_k} \ZZ_k^c \right) + \overline{\mathbf{S}}_I \ZZ_k^c \,,
    \label{eq:minsolution}
\end{equation}
where $\overline{\mathbf{S}}_I=\Id_N-\mathbf{S}_I$. In our network, we obtain $\lambda^c_k$ from $\mathbf{c}_k$ by a learned affine transformation, possibly allowing the resulting $\lambda^c_k$ to become zero. This in turn allows the network to enforce hard data-consistency.
\begin{figure}[t]
    \centering
    \includegraphics[width=\linewidth]{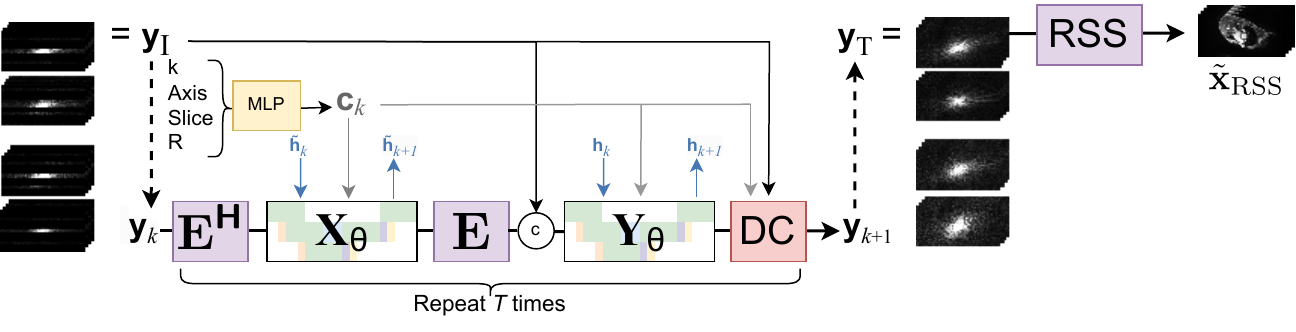}
    \vspace{-8mm}
    \caption{Schematic overview of the proposed method. It applies U-Net blocks to the multi-coil image and $k$-space data and uses intermediate (coil-wise) data-consistency blocks and a final RSS reconstruction. Note the information sharing between different iterations via the hidden latent variables $h_k$ ($k$-space) and $\tilde{h}_k$ (image-space). Both U-Nets are conditioned on $\mathbf{c}_k \in \R^{192}$, in which we encode information about the axis, slice, iteration, and acceleration factor via a MLP (details in the main text). In each iteration, the DC block performs \eqref{eq:minsolution} with $\lambda^c_k$ obtained from $\mathbf{c}_k$ by a learned affine mapping.}
    \label{fig:approach}
\end{figure}

\begin{figure}[t]
\centering
\includegraphics[width=\linewidth]{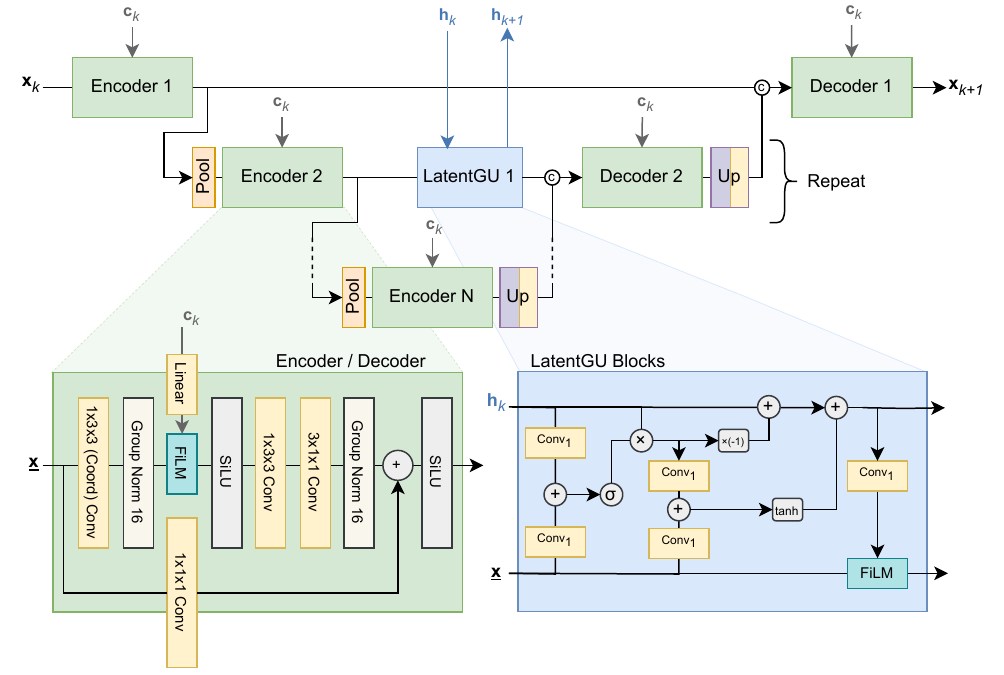}
\vspace{-8mm}
    \caption{Detailed view of one of the U-Nets used within our reconstruction network as $X_\theta$ (at 4 resolution scales) and $Y_\theta$ (3 resolution scales). The residual blocks employed as encoder/decoder apply FiLM conditioning \cite{film} and separate spatial/temporal convolutions. We introduced MGU-inspired \cite{mgu} \textit{LatentGU} blocks in the skip connections at different resolutions for sharing information between different iterations.}
    \label{fig:net}
\end{figure}

\begin{figure}
    \begin{tabular}{>{\centering}m{.24\linewidth}>{\centering}m{.24\linewidth}>{\centering}m{.24\linewidth}>{\centering}m{.24\linewidth}}
             Coilwise U-Net & E2E-VarNet & Proposed & GT / RSS
        \end{tabular}
    \resizebox{\linewidth}{!}{
    \includegraphics[height=3cm]{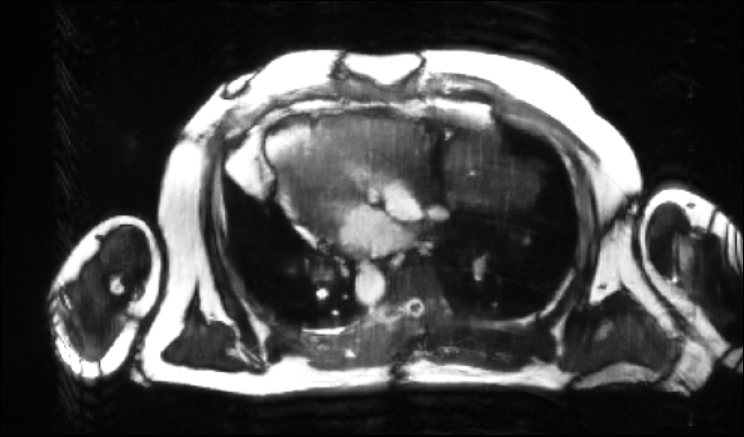}\hspace{-0.1cm}
    \includegraphics[width=1cm,height=3cm]{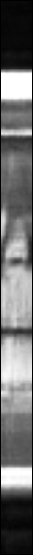}
    \includegraphics[height=3cm]{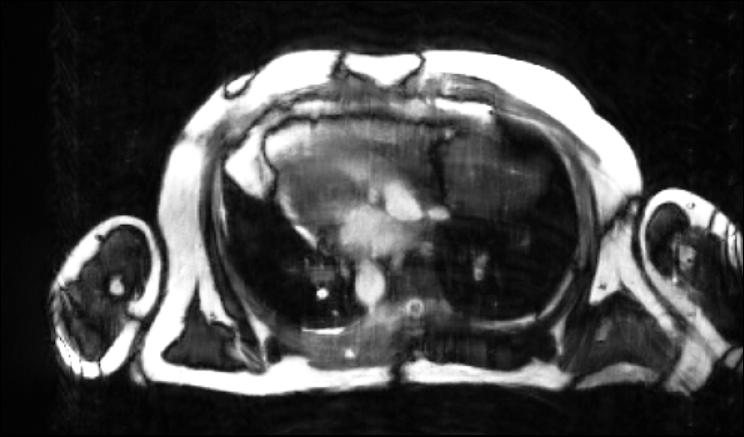}\hspace{-0.1cm}
    \includegraphics[width=1cm,height=3cm]{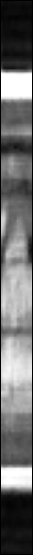}
    \includegraphics[height=3cm]{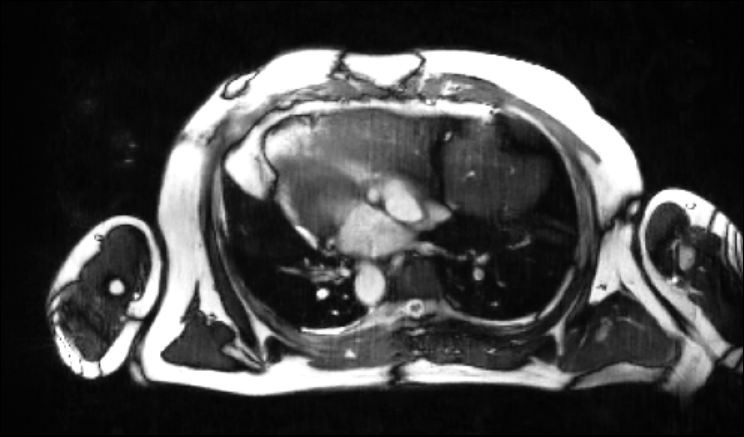}\hspace{-0.1cm}
    \includegraphics[width=1cm,height=3cm]{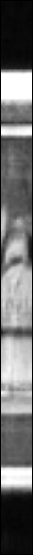}
    \includegraphics[height=3cm]{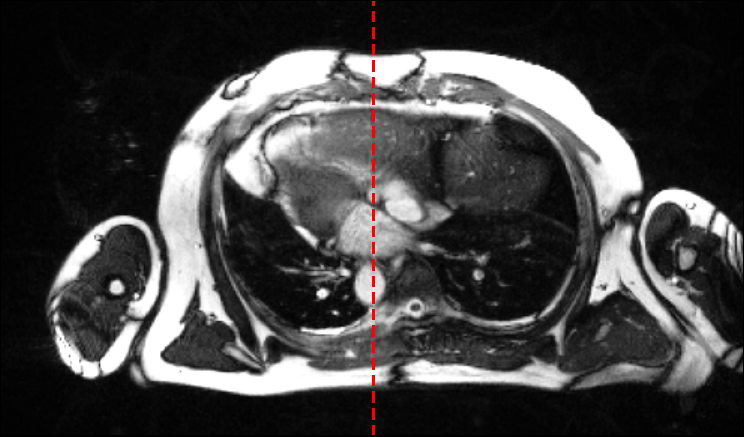}\hspace{-0.1cm}
    \includegraphics[width=1cm,height=3cm]{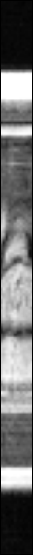}
    \hspace{-0.2cm}
    }
    \resizebox{\linewidth}{!}{
    \includegraphics[height=3cm]{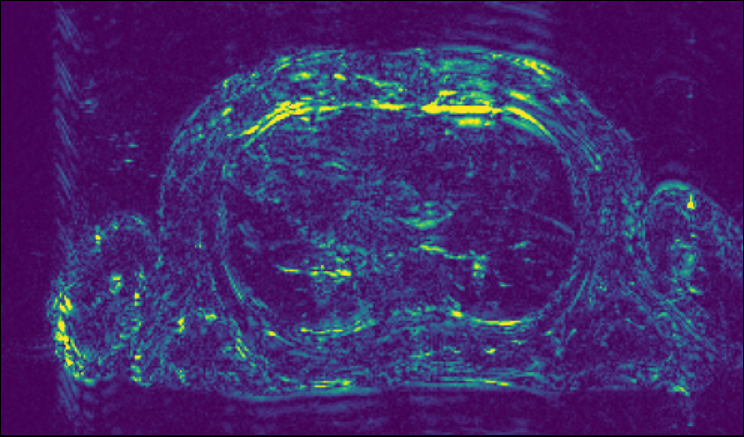}\hspace{-0.1cm}
    \includegraphics[width=1cm,height=3cm]{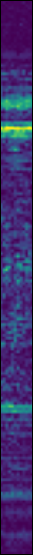}
    \includegraphics[height=3cm]{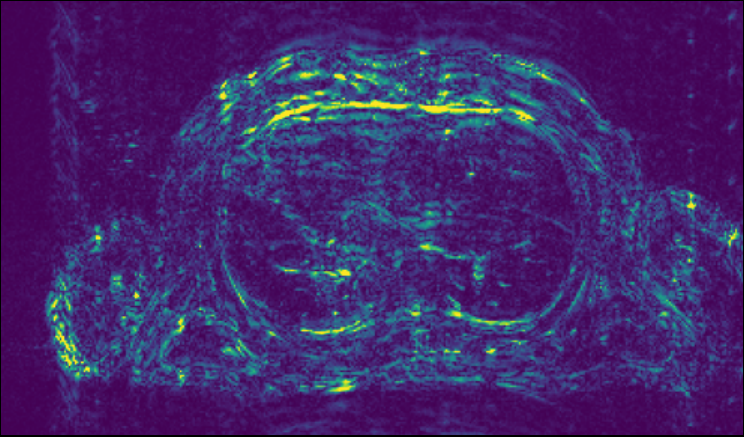}\hspace{-0.1cm}
    \includegraphics[width=1cm,height=3cm]{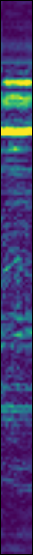}
    \includegraphics[height=3cm]{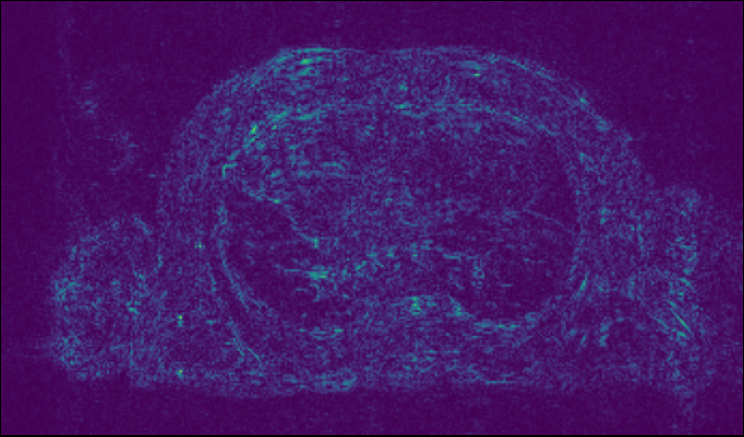}\hspace{-0.1cm}
    \includegraphics[width=1cm,height=3cm]{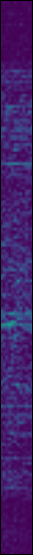}
    \includegraphics[height=3cm]{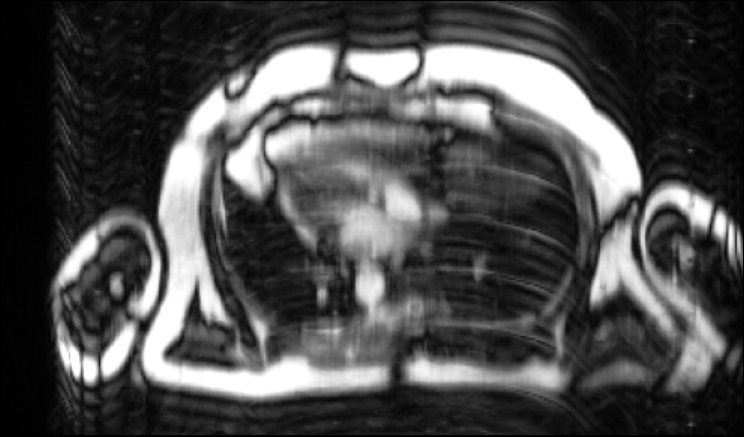}\hspace{-0.1cm}
    \includegraphics[width=1cm,height=3cm]{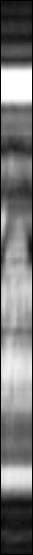}
    \hspace{-0.2cm}
    }
    \caption{An example of an LAX image reconstructed with the reported methods of comparison and our proposed approach at $R=8$ and corresponding amplified point-wise absolute error images (bottom) as well as the fully-sampled ground truth and the zero-filled RSS reconstruction. The example is in the official \textit{train} split of the CMRxRecon challenge, but has not been used for training.}
    \label{fig:results}
\end{figure}

Figure \ref{fig:approach} shows a schematic overview of our proposed approach.  
The U-Net blocks $X_\theta$ and $Y_\theta$ within our network are depicted in Figure \ref{fig:net}. These blocks include encoders/decoders comprising separate SiLU-activated spatial convolutions in the (xy)-dimensions and time-wise convolutions \cite{qiu2017learning,zimmermann2023pinqi}. We employ group normalization, FiLM-based conditioning \cite{film,nichol2021} on $\mathbf{c}_k$, and, in the encoders, CoordConvs \cite{coordconv}. Notably, we introduce novel \textit{LatentGU} blocks after each encoder except the first. These blocks, inspired by the minimal gated unit \cite{mgu}, consist of $1\times1\times1$ convolutions. They are used to FiLM-modulate the encoded features, depending on the incoming hidden state $\mathbf{h}_k$. Additionally, they form  $\mathbf{h}_{k+1}$ for the next iteration based on the current encoded features to share information across the $T$ iterations.
The MLP used to obtain the 192-dimensional $\mathbf{c}_k$ from the iteration counter $k$, the one-hot encoded imaging axis, slice position, and acceleration factor consists of two SiLU-activated hidden layers with 192 features each. Overall, our network has 7\, million trainable parameters.

 We used the AdamW optimizer \cite{adamw} with a cosine annealed learning rate and linear warmup \cite{lrschedule} and gradient accumulation over 4 samples The training was performed slice/view-wise on the full training dataset provided by the challenge except for one subject which we held back and used as an additional validation sample for hyper-parameter tuning. As data augmentation techniques, we employed spatial and temporal flips as well as shuffling of the different coils. We used oversampling of the fewer LAX views compared to the SAX slices.
 
 The training objective $\mathcal{L}(\theta)$ was a linear combination of coil-wise $L_1$-loss, as well as SSIM- and $L_2$-loss between the final RSS and the reference RSS-reconstruction. Due to the particular evaluation metrics used in the challenge, we included an additional term penalizing a deviation of the intensity of the brightest pixel in each slice at each timepoint from the corresponding ground-truth RSS's brightest pixel,
 \begin{align}
 \mathcal{L}(\theta)&= \frac{1}{N} \left(\frac{1}{N_c}\|\Ed^\herm \YY-\Ed^\herm \tilde{\YY}\|_1
 -\alpha_{1}\mathrm{SSIM}(\XX_{\mathrm{RSS}},\tilde{\XX}_{\mathrm{RSS}}) +
 \alpha_2  \|\XX_{\mathrm{RSS}}- \tilde{\XX}_{\mathrm{RSS}}\|_2^2\right) \nonumber\\
 &+ \alpha_3 \|\max \XX_{\mathrm{RSS}}, \max \tilde{\XX}_{\mathrm{RSS}}\|_2^2 \,.
\end{align}
 We empirically chose $\alpha_1=1.0$, $\alpha_2=0.2$, $\alpha_3=0.05$.
 
 As methods of comparison used for guidance during the development process, we chose a basic \textit{2.5\,D} U-Net as well as the well-known end-to-end variational network \cite{sriram2020end} (E2E VarNet), which includes an explicit estimation of the CSMs and a U-Net applied to coil-combined images.
 Our implementation of the \textit{2.5D} U-Net follows the construction of \cite{zimmermann2023pinqi}
 with separate convolutions along $(xy)$-spatial dimensions and temporal dimension  \cite{qiu2017learning} with 3.5\,million trainable parameters. The input of the U-Net consists of a concatenation of real and imaginary parts of each coil image and the RSS reconstruction. The output is combined with a residual for each coil before an RSS reconstruction is performed.
Our implementation of E2E VarNet follows the one in the original work \cite{sriram2020end} and extends it for the application to cardiac cine MR data. Based on preliminary experiments, we also employed 2.5\,D U-Nets instead of computationally more demanding 3\,D U-Nets \cite{Hauptmann2019} for the image- and CSMs-refinement modules. This approach has a total of 8 million trainable parameters.
Both methods were trained with AdamW, a cosine learning rate schedule with warmup, and a combination of $L_1$, $L_2$, and SSIM as loss functions with weightings and maximum learning rate determined by a grid search. An implementation of the proposed method with the final hyperparameters used in our challenge submission, trained weights, and all training code will be made available at \url{https://github.com/fzimmermann89/CMRxRecon}.

\subsection{Ablations}
\vspace{-1.5mm}
We performed the following ablations to investigate the importance of specific parts our our proposed reconstruction network.

\textit{K-Space U-Net:}
We removed the $k$-space U-Net \eqref{eq:kspace_cnn} from the network, thus leaving only the learned image space refinement and data-consistency, to investigate the influence of $Y_\theta$.

\textit{Image-Space U-Net:}
Similarly, to investigate the influence of the image-space U-Net, $X_\theta$ \eqref{eq:img_cnn}, we performed an experiment with only the learned image space refinement and data-consistency layer in each iteration of the cascade.

\textit{LatentGU Blocks:}
We removed the LatentGU blocks from both U-Nets, thus ablating the information sharing between different iterations.

\textit{Conditioning on the Meta-Information:} 
We removed all conditioning on meta-information such as acceleration factor, axis, and current iteration $k$ embedding in $\mathbf{c}_k$. Specifically, we remove the MLP as well as the FiLM blocks inside the U-Nets. Also, we replaced the mapping from $\CC_k$ to $\lambda^c_k$ inside of the DC layer by $T\cdot N_c$ learned constant parameters.

\textit{Iterations of the Cascade:} 
Finally, we set $T=1$, thus performing only a single iteration of \eqref{eq:img_cnn}, \eqref{eq:kspace_cnn} and \eqref{eq:scheme_min_problem} each to investigate whether the cascading approach is beneficial in our specific network.

\subsection{Extension to Quantitative Mapping}
Although we mainly focus on cine imaging, our proposed architecture can also be applied to the CMRxRecon mapping track. Similarly to the cine track, it involves reconstructing qualitative images for evaluation. Thus, approaches that integrate image reconstruction and the parameter estimation task \cite{zimmermann2023pinqi} within a single reconstruction network are not readily applicable. However, our \textit{NoSENSE} approach can easily be adapted. The main difference between these two tasks in the CMRxRecon challenge lies in the effect of the time-dimension. Whereas in the cine task, between neighboring time points the images change mainly in anatomy due to cardiac motion, in the mapping track the changes are mainly in the form of changes in the contrast due to the signal preparation \cite{yang2022disq}. We slightly modify our architecture by including a learned data-dependent normalization before each image-space U-Net \cite{spade}. Here, for each pixel, the RSS intensities at time points are fed as channels into a simple two-layer, 128 feature, window size 3 CNN, which outputs two scaling factors for each time point for each pixel to be applied before and after each image space U-Net, respectively.
\section{Results}
We report the results of the CMRxRecon cine reconstruction, T1/T2-mapping reconstruction task, and ablation experiments for the cine task as reported by the validation leaderboard. Please note, that for all metrics, the official leaderboard normalizes the ground truth by its maximum value and the prediction by its maximum value, respectively.
\subsection{Cine Reconstruction}
Figure \ref{fig:results} shows an example of cine images reconstructed with the methods reported. We see that our proposed approach accurately removes undersampling artifacts and clearly outperforms baseline comparison methods. Table \ref{tab:results} lists the metrics on the (cropped) CMRxRecon validation dataset as reported by the public leaderboard.

\begin{table}[h!]
\caption{Results of our method compared to RSS reconstruction, a coil-wise U-Net baseline, and an E2E-VarNet with explicit sensitivity map estimation as reported by the CMRxRecon validation leaderboard.}
\centering
\vspace{-2mm}
{\footnotesize{
\begin{tabular}{L{0.5cm} L{1.2cm} L{1.2cm}| C{1.8cm}|C{1.8cm}|C{1.8cm}|C{1.8cm}}
\hline \hline
&&              & \textbf{RSS}  & \textbf{U-Net (Coilwise)} & \textbf{E2E VarNet} & \textbf{Proposed}  \\
\hline
\multirow{9}{*}{\rotatebox[origin=c]{90}{\textbf{Multi-Coil LAX}}}&& \textbf{PSNR} & 22.75 &  29.14 & 30.60 & \textbf{36.43} \\
&$R=4$ & \textbf{NMSE} & 0.166 &  0.030 & 0.025 & \textbf{0.006} \\
&& \textbf{SSIM} & 0.640 &  0.828 & 0.862 & \textbf{0.942} \\
\cline{2-7} 
&& \textbf{PSNR} & 22.90 &  26.96 & 28.24 & \textbf{32.08} \\
&$R=8$ & \textbf{NMSE} & 0.168 &  0.052 & 0.041 & \textbf{0.015} \\
&& \textbf{SSIM} & 0.639 &  0.785 & 0.816 & \textbf{0.891} \\
\cline{2-7} 
&& \textbf{PSNR} & 22.77 & 26.47 & 27.67 & \textbf{31.30} \\
&$R=10$ & \textbf{NMSE} & 0.175 &  0.059 & 0.045 & \textbf{0.018} \\
&& \textbf{SSIM} & 0.634 &  0.773 & 0.805 & \textbf{0.880} \\
\hline
\multirow{9}{*}{\rotatebox[origin=c]{90}{\textbf{Multi-Coil SAX}}}&& \textbf{PSNR} & 24.43 &  32.69 & 33.35 & \textbf{39.92} \\
&$R=4$ & \textbf{NMSE} & 0.136 &  0.016 & 0.018 & \textbf{0.003} \\
&& \textbf{SSIM} & 0.718 &  0.888 & 0.907 & \textbf{0.965} \\
\cline{2-7}
&& \textbf{PSNR} & 23.70 &  30.22 & 30.90 & \textbf{35.42} \\
&$R=8$ & \textbf{NMSE} & 0.163 &  0.028 & 0.027 & \textbf{0.008} \\
&& \textbf{SSIM} & 0.692 &  0.848 & 0.869 & \textbf{0.929} \\
\cline{2-7}
&& \textbf{PSNR} & 23.40 &  29.50 & 30.15 & \textbf{34.22} \\
&$R=10$ & \textbf{NMSE} & 0.174 &  0.034 & 0.031 & \textbf{0.011} \\
&& \textbf{SSIM} & 0.685 &  0.835 & 0.856 & \textbf{0.915} \\
\hline \hline
\end{tabular}    
\label{tab:results}
}
}
\end{table}

\begin{table}[h!]
\caption{Results of the ablation experiments on the CMRxRecon cine validation leaderboard, highlighting the importance of the different parts of your proposed network. Shown are average values over different acceleration factors.}
\label{tab:ablation}
\centering
\vspace{-2mm}
\begin{tabular}{L{4cm}| C{1.8cm}|C{1.8cm}|C{1.8cm}|C{1.8cm}}
\hline \hline
\multirow{2}{*}{\textbf{Ablation}}& \multicolumn{2}{c}{\textbf{Multi-Coil SAX}} & \multicolumn{2}{c}{\textbf{Multi-Coil LAX}} \\
&\textbf{PSNR} & \textbf{SSIM} & \textbf{PSNR} & \textbf{SSIM} \\ 
\hline
No $k$-space U-Net       & 34.88 &  0.921 & 31.87  & 0.887\\ 
No Image-space U-Net     & 28.62 &	0.811 &	26.24 &	0.760\\ 
No LatentGU              & 35.11 &	0.923 &	32.06 &	0.888\\ 
No Conditioning          & 34.64 &	0.916 & 31.58 & 0.872\\ 
Single Iteration ($T$=1) & 33.69 &	0.902 &	30.69 &	0.860\\ 
\hline
Full Proposed Network    & \textbf{36.46} & \textbf{0.936} & \textbf{33.22} & \textbf{0.904} \\
\hline \hline
\end{tabular}
\end{table}

\subsection{Ablations}
In Table \ref{tab:ablation} we report the results of the different ablations, evaluated on the (cropped) cine validation data of the CMRxRecon challenge in terms of SSIM and PSNR compared to our full network as proposed.

\subsection{Quantitative Mapping Image Reconstruction}
Application of our network trained on the cine task to the reconstruction of the images used in parameter mapping without any additional training resulted in $0.900$\,SSIM and $32.35$\,PSNR on the CMRxRecon mapping track validation leaderboard.  Using the same architecture and re-training on the mapping task training dataset resulted in $0.934$\,SSIM / $34.92$\,PSNR. By incorporating the learned data-dependent normalization, this improved further to $0.942$\,SSIM / $35.57$\,PSNR. As a comparison, a simple zero-filled RSS reconstruction achieves $0.701$\,SSIM / $23.06$\,PSNR on the same validation dataset.


\section{Discussion and Conclusion}
We have presented a novel approach for the reconstruction of undersampled $k$-space cardiac MR data inspired by unrolled dual-domain networks \cite{adler2018learned,kiki} with recurrent blocks \cite{yiasemis2022recurrent} and adaption via feature modulation \cite{film}. A fundamental aspect of our method is its direct estimation of multi-coil $k$-space data as opposed to the coil-combined image, as well as the usage of coil-wise image data within the convolutional blocks. Therefore, it is custom-tailored to the specific CMRxRecon challenge task. This design choice was also influenced in part by the unexpectedly strong performance of a single U-Net when applied to non-coil-combined images in contrast to the typically more potent E2E VarNet \cite{sriram2020end}.

A performance gap between LAX and SAX is evident in Table \ref{tab:results}, which might be attributed to our unsophisticated oversampling approach not completely mitigating the impact of the fewer LAX images present in the training dataset. So far, we trained the network only in a supervised manner. However, the challenge organizers have provided a relatively large validation dataset across various acceleration factors. While this dataset lacks target fully sampled RSS-reconstructions, it can be harnessed for training using self-supervised learning techniques \cite{yaman2020self}. Also, the hyperparameters of the loss and the U-Net blocks might be further optimized outside the time constraint of a challenge. 

In our coarse ablations, we were able to show the great importance of the image-space U-Net and the incremental improvements by including a $k$-space U-Net, the latent space information sharing between different iterations, and the conditioning on the meta-information.  Finally, multiple iterations, i.e., $T>1$, are required for good performance.
A more detailed investigation of the effect of the different parts of our architecture will be explored in future works.

An inherent limitation of the proposed approach is its substantial GPU memory requirements, even when compared to other unrolled methods, especially during training. This is due to the increased number of features necessary for the highest resolution within the U-Nets. We successfully mitigated the issue by activation checkpointing of these less computationally dense but memory-intensive blocks \cite{checkpointing}, enabling single GPU training.

The straight transfer of the network trained on cine data to quantitative mapping data already resulted in surprisingly good performance. By proper re-training and incorporating a learned normalization layer we further improved the result. A more sophisticated normalization approach can be further explored in future works. 

\subsubsection*{Acknowledgments}
This project has received funding from the European Partnership on Metrology, co-financed from the European Union’s Horizon Europe Research and Innovation Programme, and by the Participating States. This work was supported in part by the Metrology for Artificial Intelligence in Medicine (M4AIM) project, which is funded by the German Federal Ministry for Economic Affairs and Climate Action (BMWi) as part of the QIDigital initiative.
\newpage
\bibliography{references}
\bibliographystyle{splncs04}

\newpage
\section*{Supplement}
\begin{table}[h!]
\caption{CMRxRecon Challenge Participation Information}
\centering
\begin{tabularx}{\textwidth}{L{5cm}|X}
\hline \hline
Challenge Track & Cine and Mapping\\
University$/$organization                                               & Physikalisch Technische Bundesanstalt (PTB), Braunschweig \& Berlin, Germany\\
Single-channel or multi-channel                                         & Multi-Coil\\
Hardware configuration               & Nvidia A6000 (48GB), Intel Xeon 6326 (16 Core)\\
Training time                                                           & approx. 48h (Cine), 12h (Mapping) \\
Inference time                                                          & approx. 1 hour (full validation set)\\
Validation Performance (Cine)                    & PSNR 34.89, SSIM 0.920, NMSE 0.010\\ 
Validation Performance (Mapping)                     &PSNR 35.57,
SSIM 0.942, NMSE 0.008\\
Docker submitted                                                       & Yes\\
Use of segmentation labels                                              & No\\
Use of pre-training                   & No, not allowed by the challenge\\
Data augmentation                                                       &  Spatial and temporal flips, shuffling of the different coils\\
Data standardization                                                    & Mean 0, Standard Deviation 1 \\
Model parameter number                                                  & approx. 7 million trainable parameters \\
Loss function                                                           & Combination of coil-wise $L_1$-loss, SSIM and $L_2$-loss\\
Incorporation of a physical model    & Yes, frame-wise 2D FFT\\
Use of unrolling                                                        & Yes \\
Application of k-space fidelity                                         & Yes, approximate data-fidelity as a solution of a minimization problem\\
Model backbone                                                          & 2.5 D U-Nets with LatentGU and FilM, MLP\\
Complex-valued operations                                  &  complex-valued images represented by real/imag. channels in U-Nets\\
\hline \hline
\end{tabularx}
\end{table}

\end{document}